# Large-area, high-NA Multi-level Diffractive Lens via inverse design

Monjurul Meem,[1] Sourangsu Banerji,[1] Christian Pies,[2] Timo Oberbiermann,[2] Berardi Sensale-Rodriguez[1] and Rajesh Menon[1,3,*]

[1]*Department of Electrical & Computer Engineering, University of Utah, Salt Lake City UT 84112 USA*
[2]*Heidelberg Instruments Mikrotechnik, Heidelberg Germany.*
[3]*Oblate Optics, Inc. San Diego CA 92130 USA*

*Corresponding author:* rmenon@eng.utah.edu

**Abstract:** Flat lenses enable thinner, lighter, and simpler imaging systems. However, large-area and high-NA flat lenses have been elusive due to computational and fabrication challenges. Here, we applied inverse design to create a multi-level diffractive lens (MDL) with thickness <1.35µm, diameter of 4.13mm, NA=0.9 at wavelength of 850nm. Since the MDL is created in polymer, it can be cost-effectively replicated via imprint lithography.

Large-area metalenses are challenging due to their deep sub-wavelength critical dimensions, relatively large aspect ratios, and the need for high-refractive-index materials [1]. Therefore, only low-NA, large-area [1] or high-NA, small-area metalenses [2] have been demonstrated. Furthermore, recent work has also indicated that the unit-cell design methodology used in the vast majority of metalenses places an upper bound on focusing efficiencies at high numerical aperture (NA) [3]. Unit-cell design approach is critical to design large-area metalenses due to the huge computational cost of alternatives. Therefore, large-area, high-NA metalenses are extremely challenging.

On the other hand, binary-diffractive lenses at high NA have been proposed, [4,5] and demonstrated in air [6] and also under water immersion for NA>1 [7]. However, these suffer from relatively low efficiencies. Previously, we showed that high efficiency at all NAs could be achieved by exploiting inverse-design in conjunction with 2.5D microstructures [8]. Specifically, we showed that such multi-level diffractive lenses (MDLs) offer the same or better performance when compared to metalenses, but with much simpler fabrication. Relatively large area (diameter=15.2mm) MDLs have also been demonstrated experimentally [9]. The concept of inverse design in optics, where the geometry of the device is computed based upon the desired photonic functionality has been applied for many years in light-trapping, [10] in photovoltaics, [11] in integrated photonics, [12] in free-space metasurfaces [13] and in MDLs [14]. In this memorandum, we apply inverse design and experimentally demonstrate an MDL with NA=0.9, diameter=4.13mm, focal length=1mm and operating wavelength, $\lambda$=850nm. Most importantly, we also show that high efficiency at high-NA is indeed possible with appropriate design of the MDL, something that has been considered very challenging for all flat lenses [15].

Our MDL is comprised of 4589 concentric rings, each of width 450nm and heights varying between 0 and 1.35µm. The heights were quantized to at most 100 discrete steps. The MDL was patterned in a positive-tone photoresist (maP1200G, Microresist Technology GmbH). The height distribution was chosen based on our inverse-design algorithm [8, 16]. The MDL was patterned using grayscale-optical lithography via a laser-pattern generator (DWL 66+, Heidelberg Instruments Mikrotechnik GmbH) [16]. Micrographs of the fabricated MDL are shown in Fig. 1.

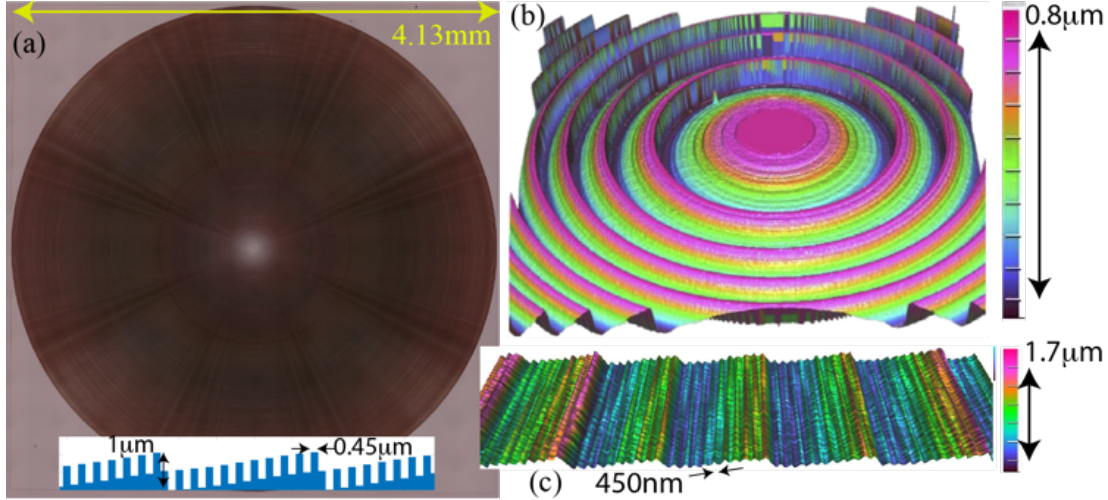

*Fig. 1. Details of the fabricated MDL. (a) Widefield transmission optical micrograph. Inset shows radial cross-section through the last 50 rings of the design. The full design is in Fig. S1 [16]. Scanning-confocal images of (b) the central rings and (c) a portion of the outer rings.*

We characterized our device by illuminating it with a collimated beam at λ=850nm (bandwidth = 15nm, superK with Varia filter, NKT Photonics) and captured the point-spread function (PSF) by first magnifying it by 230 X and then recording the magnified PSF on a monochrome CMOS image sensor (DMM 27UP031-ML The Imaging Source, pixel size = 2.2μm) (see details in [16]). The recorded PSF and a cross-section through its center are shown in Fig. 2(a), where the full-width at half-maximum (FWHM) is confirmed to be 560nm. The diffraction-limited FWHM is 472nm. It is interesting to note the absence of any sidelobes, which are present even in well corrected high-NA microscope objectives. We calculated the modulation-transfer function from the measured PSF and plotted it in Fig. 2b (inset shows the same data in log scale). At 10% contrast, the resolution is >300lp/mm.

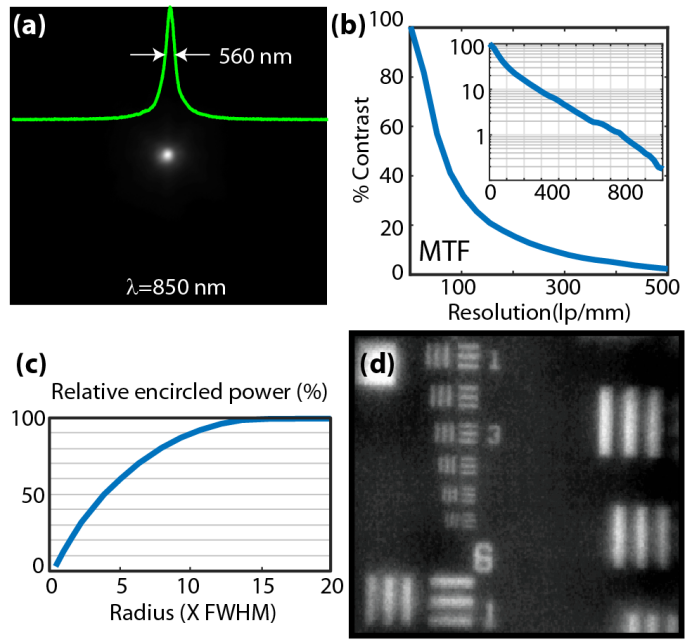

*Fig. 2. Characterization of the MDL. (a) Measured PSF. (b) MTF obtained from the PSF (Inset: contrast in log-scale). (c) Relative encircled power vs radius of spot. (d) Image of resolution chart. Illumination for (a) was a laser, while that for (d) was an LED flashlight, both centered at λ=850nm [16].*

There have been significant inconsistencies in the reporting of experimental efficiencies of flat lenses. In most cases, the focusing efficiency is defined as the ratio of optical power inside the focused spot to the power incident on the lens. However, the area of the focused spot is seldom consistently defined. This area is sometimes defined as a circle with a diameter of 3 X FWHM, [17], or 8 X FWHM, [18] and often, the area is not reported at all [19]. This last reference is particularly troubling because the claimed focusing efficiency of 86% at NA=0.9 has been repeated in various review articles [6], but it is far higher than what is theoretically predicted to be possible in a recent article [3]. According to [3], the upper bound on efficiency for NA=0.9 metalens (using the unit-cell design method) is 32%. In order to avoid these pitfalls, instead of a single efficiency number, we prefer to plot the experimental relative encircled power as a function of the radius from the center of the focused spot in Fig. 2(c). The relative encircled power is defined as the ratio of the optical power within a spot centered on the optical axis of a given radius to the total power within the aperture of the lens. This is a much more common metric used to characterize conventional refractive lenses [20]. The results indicate that more than 90% of the power is confined within a spot of radius ~ 10 X FWHM. We also emphasize that since our materials have a low refractive index (n=1.62 at λ=850nm), the transmission efficiency is very high (~90%, see [16]).

Finally, we also captured an image of the AirForce resolution chart, which is reproduced in Fig. 2(d), where one can see that all lines are well resolved, which is consistent with the measured PSF. The distance between the target and the MDL was 1.5mm, while the distance between the MDL and the image plane was 3mm [16]. The illumination was from an 850nm-LED flashlight (bandwidth=35nm, see Fig. S8).

In conclusion, we demonstrated that inverse design coupled with high-resolution grayscale lithography can achieve high-NA, large area diffractive lenses. Since these lenses are made in polymer material, they can be manufactured at low cost via high-volume replication methods such as imprint lithography.

**Funding:** National Science Foundation (NSF) (1351389, 1828480, 1936729); Office of Naval Research N66001-10-1-4065.

**Acknowledgment**. We thank Dr. Brian van Devener for the use of a high-NA objective for characterization.

**Disclosures.** RM is co-founder of Oblate Optics, Inc.

# Supplementary Information

## 1. Details of Design

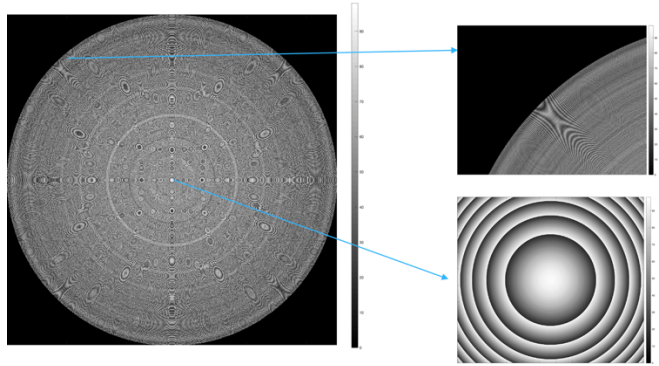

Fig. S1: The design has a total of 4589 rings. Each ring is 0.45µm wide, while its height varies between 0 and 1.35µm.

## 2. Fabrication

For the grayscale lithography step, glass substrates were coated with a low-contrast positive-tone photoresist. The desired topography is obtained by illuminating the resist with a spatially modulated light intensity and subsequent selective removal in a developer solution (fig. S2).

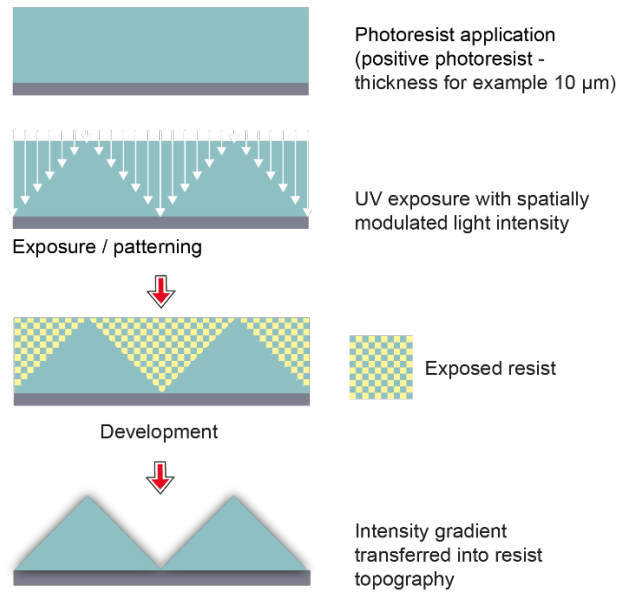

Fig S2: Process flow of grayscale lithography.

The DWL 66+ used for this exposure was equipped with a diode laser of 405 nm wavelength and provides a 50 nm pixel grid. The optical engine of the system allows to expose each pixel independently with one of 65000 intensity levels, which is defined by the pixel's grayvalue.

The MDL layout was defined in a bitmap image, where the grayvalue of each pixel represents the desired height level. The pattern was composed of 100 equidistant height levels, corresponding to a step size of 13.5 nm. (This paragraph might also be shifted to section 1)

To take into account the nonlinear behavior of the photoresist as well as proximity effects between neighboring pixels, a pixel-wise grayvalue correction was applied to the pattern. For this purpose, a reference pattern pattern was exposed, developed and measured with a confocal microscope first to determine the contrast curve (remaining resist height as a function of exposure intensity) of the resist. This measurement result, the lens pattern, the target profile (defining the target height of each grayvalue), and system parameters of the DWL 66+ served as input for the 3D-PEC module of the software BEAMER (GenISys GmbH), which calculates a modified pattern image based on a numerical simulation of the exposure process. In order to improve the precision of the correction, the number of grayvalues was increased to 1000 in this process.

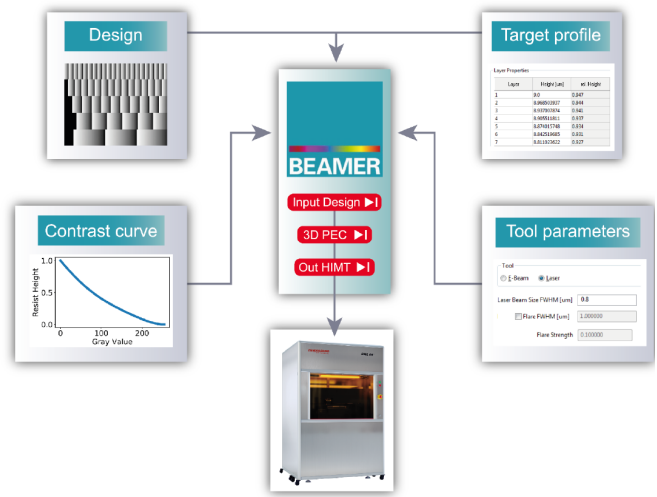

Fig. S3: Process flow of the 3D-PEC correction

### 3. Focal spot characterization

The MDL was illuminated with expanded and collimated beam from a SuperK EXTREME EXW-6 source (NKT Photonics). The wavelength and bandwidth was tuned with SuperK SELECT filter (NKT Photonics) for near infrared wavelengths (800nm-1400nm). The focal planes of the flat lenses were magnified using a 150x objective (Reichert Plan Apo) with the NA of 0.95 and tube lens (ITL200, Thorlabs) and imaged onto monochrome sensor (DMM 27UP031-ML, Imaging Source). The exposure time of the sensor was carefully adjusted to avoid pixel saturation. A dark frame was also captured with all the light source turned off. *The* gap between objective and tube lens was ~120 mm and that between the sensor and the backside of tube lens was about 180mm. The magnification of the microscope was found by imaging a calibration sample with known feature dimensions.

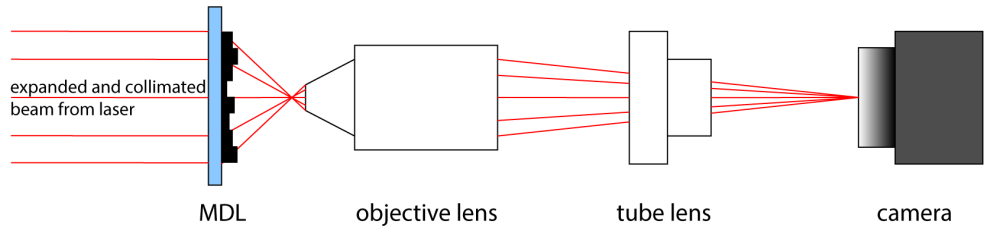

Fig. S4: Schematic of the set-up used to capture the focal spots of the MDL. An expanded and collimated laser beam is focused by the MDL. An objective lens paired with a tube lens is used to form a magnified image of the focal spot on a CMOS camera.

## 4. Imaging Setup

For imaging experiment, we used a1951 USAF resolution test chart (R3L3S1N, Thorlabs) as object. The target was illuminated with a 850nm IR flash light (5W 4chip 850nm, Make The One). A diffuser was also placed behind the object to make the illumination more uniform. The image formed by MDL was captured using a monochrome sensor (DMM 27UP031-ML, The Imaging Source). The exposure time was adjusted to ensure that the images did not get saturated. In each case, a dark frame was recorded and subtracted from the USAF target images.

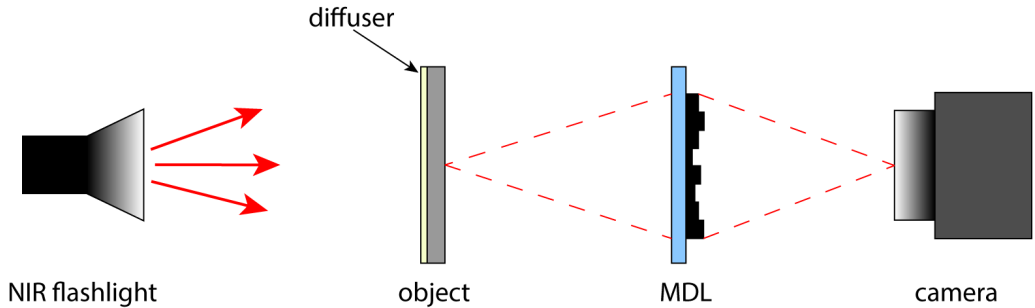

Fig S5: Schematic of the imaging setup. The object was illuminated with a 850nm IR flash light and the corresponding image formed by the MDL was captured on a monochrome camera.

## 5. Measured Focal Spots

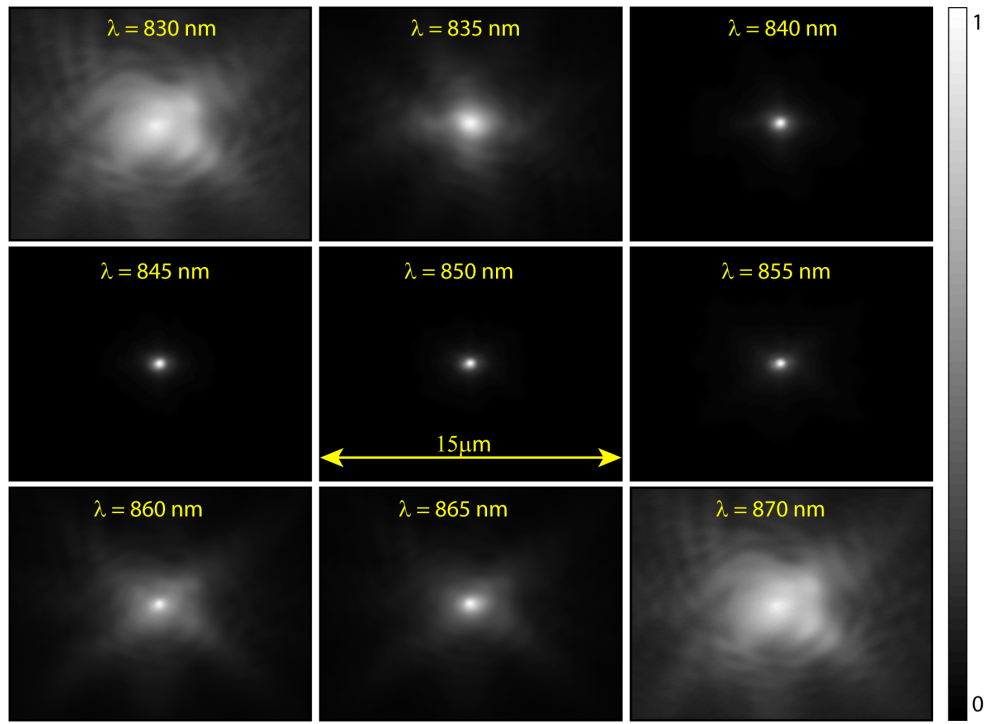

Fig S6: Measured focal spot of the MDL at different wavelengths. The MDL was illuminated with a collimated and expanded laser. The wavelength of illumination was tuned from 830nm to 870nm in a step of 5nm. An objective lens paired with a tube lens is used to form a magnified image of the focal spot on a CMOS camera.

## 6. Spectrum measurements

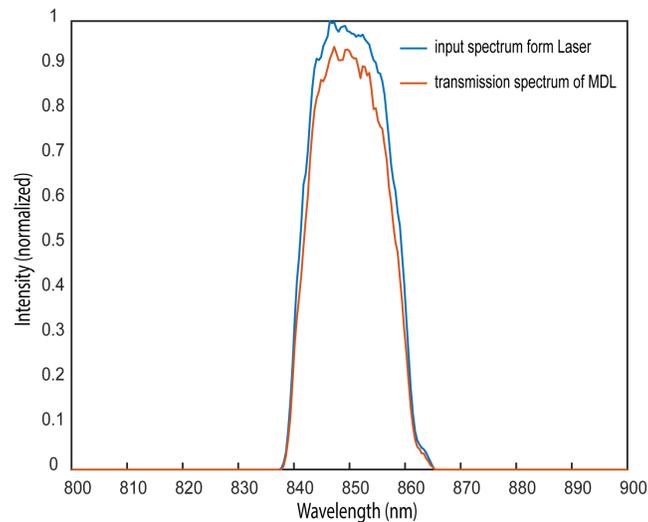

Fig S7: Transmission spectrum of MDL. The input spectrum of laser is also provided for reference.

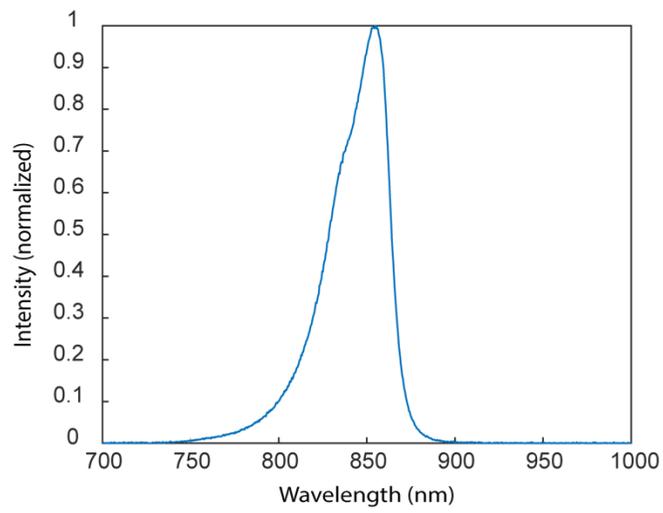
Fig S8: Spectrum of 850nm-LED flashlight.